\documentclass[12pt]{article}
\usepackage{graphicx}
\usepackage{bm}
\usepackage{balance}
\usepackage{color,soul}
\usepackage[cmex10]{amsmath}
\usepackage{amsfonts}
\usepackage{gensymb}
\usepackage{esvect}
\usepackage{caption}
\usepackage{float}
\usepackage{wrapfig}
\usepackage{epsfig}
\usepackage{psfrag}
\usepackage{amssymb}
\begin{document}
\baselineskip 12pt

\begin{center}
\textbf{\Large Low Complexity Symbol-Level Design for Linear Precoding Systems} \\
\vspace{5mm}
\begin{tabular}{cc}Jevgenij Krivochiza &  Ashkan Kalantari\\[1.0\baselineskip]
Symeon Chatzinotas & Bj{\"{o}}rn Ottersten\\
\multicolumn{2}{c}{University of Luxembourg} \\
\multicolumn{2}{c}{Interdisciplinary Centre for Security, Reliability and Trust} \\
\multicolumn{2}{c}{Luxembourg, Luxembourg city, L-1855} \\[1.2\baselineskip]
\end{tabular}  \verb+{jevgenij.krivochiza, ashkan.kalantari}@uni.lu+ \\
\verb+{symeon.chatzinotas, bjorn.ottersten}@uni.lu+ \end{center}

\newcommand{\tr}{\mathrm{tr}}
\newcommand{\tx}{\mathrm{Tx}}
\newcommand{\rx}{\mathrm{Rx}}
\newcommand{\tsup}[1]{\ensuremath{\mathsf{#1}}}

\begin{abstract}
The practical utilization of the symbol-level precoding in MIMO systems is challenging since the implementation of the sophisticated optimization algorithms must be done with reasonable computational resources. In the real implementation of MIMO precoding systems, the processing time for each set of symbols is a crucial parameter, especially in the high-throughput mode. In this work, a symbol-level optimization algorithm with reduced complexity is devised. Performance of a symbol-level precoder is shown to improve in terms of the processing times per set of symbols. 
\end{abstract}

\section{Introduction}  
Exploiting the interference components in linear precoding systems results in total transmit power reduction compared to conventional precoding technique such as using Zero-Forcing (ZF) or Minimum Mean Square Error (MMSE) \cite{DBLP:journals/corr/abs-1303-7454}, \cite{7417066}, \cite{7091022}. Symbol-level precoding to create constructive interference between the transmitted symbols has been developed using advanced optimization frameworks. However, practical implementation of precoded multiple-input multiple-output system (MIMO) is challenging. The computantional complexity of such optimal precoders grows as number of possible symbol positions in multi-level constellations increases. The symbol-level precoding implemented in previous works use Non-Negative Least Squares (NNLS) algorithm to design the optimal precoder only in scenarios of PSK constellations. In cases of multi-level constellations the design problem falls back to standard convex optimization, which results in non-trivial solution \cite{7544467}, \cite{7472324}, \cite{AlodehCO16}.

In this work, we focus on a novel symbol-level precoding design, which does not exceed the complexity of NNLS problem for all symbol constellation types. The proposed technique minimizes transmit power of the precoded symbols, which are constructed using conventional linear procoder, by manipulating the phase and the amplitude of the initial unprecoded symbols. The modified symbols afterwords are precoded with conventional precoders at the transmitter. The main advantages of such approach toward design of the precoder are less processing time to design the symbol-level precoder and simplified implementation compared to other techniques. In this work, we show that the proposed algorithm requires less time to calculate the symbol-level precoder compared to other literature benchmark schemes of \cite{7544467}, \cite{7472324}, \cite{AlodehCO16}.

The rest of this paper is organized as follows. In Section \ref{optproblem}, the symbol-level power optimization for linear precoding system is derived for QPSK constellation. In Section \ref{optcondesign}, the extension of the technique is devised for optimization of M-PSK and M-APSK constellations. In Section \ref{results} benchmark results are compared and discussed. Summary of the paper results is presented in Section \ref{conclusions}.

\textit{Notation}: Upper-case and lower-case bold-faced letters are used to denote matrices and column vectors, respectively. The superscripts $(\cdot)^{H}$, $(\cdot)^{-1}$ and $(\cdot)^{T}$ represents Hermitian matrix, inverse and transpose, respectively. $\mathbf{I}_{N \times N}$ denotes $N$ by $N$ identity matrix, $\lVert \cdot \rVert_2$ is Euclidean norm or $l^2$-norm, $| \cdot |$ is absolute value and $(\circ)$ is element-wise multiplication, $\mathbf{0}$ is the all
zero vector. The real and imaginary parts of a complex value are defined as $\mathrm{Re}(\cdot)$ and $\mathrm{Im}(\cdot)$.

\section{Symbol-level optimization for conventional precoding} \label{optproblem}
\begin{wrapfigure}{R}{0.4\textwidth}
\centering
\includegraphics[trim={5cm 1cm 5cm 2cm},width=0.14\textwidth]{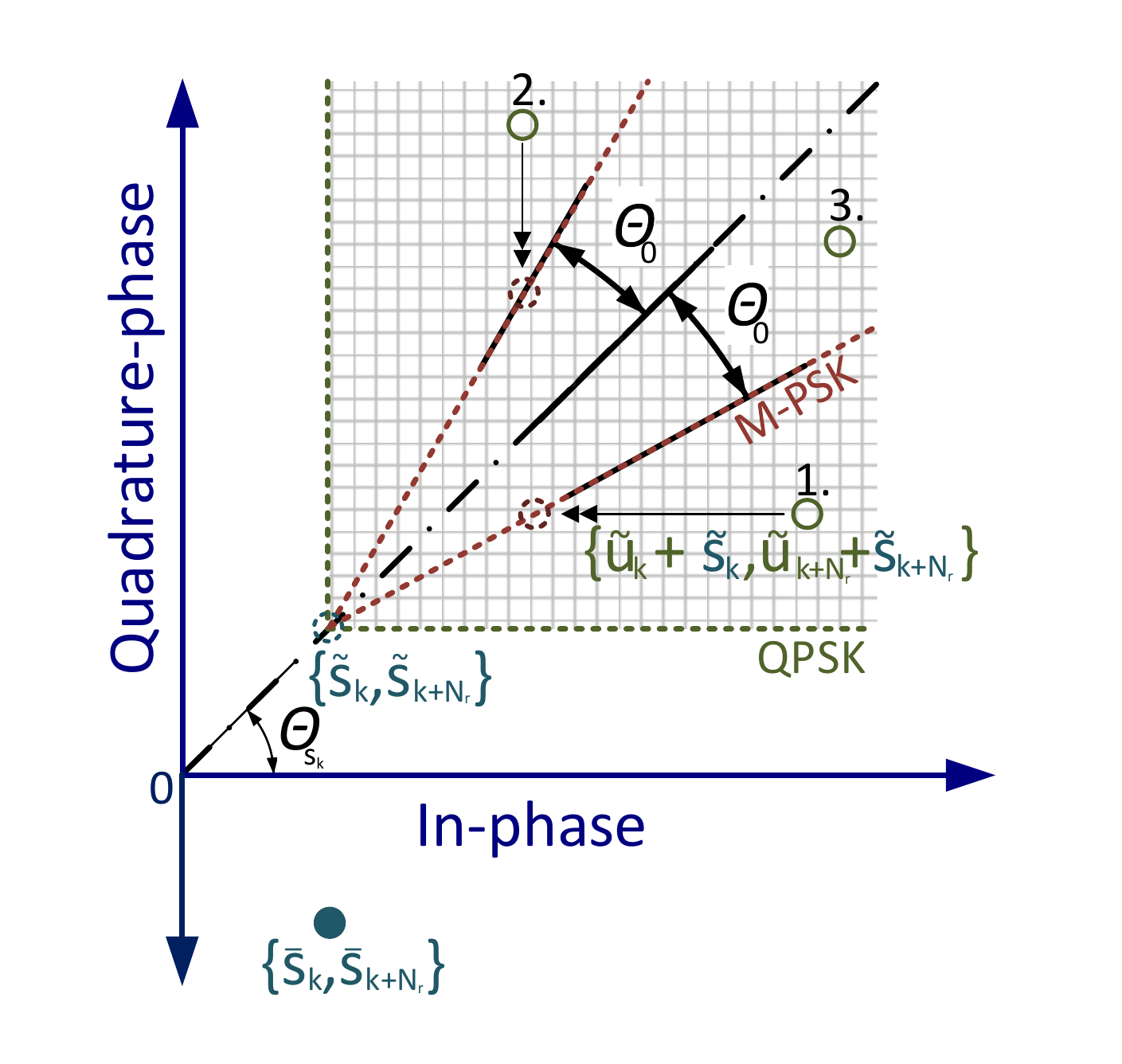}
\caption{Construction of the optimized symbol.}
\label{fig-1}
\end{wrapfigure}
We consider single-user multiple-input multiple-output system transmission system. The system consist of a transmitter with $N_\mathrm{t}$ and a receiver with $N_\mathrm{r}$ number of antennas. The received signal at the $k$-th antenna is expressed as:
\begin{equation}
y_{k} = \mathbf{h}_{k} \mathbf{x} + n_{k}. \label{chan1}
\end{equation}
where $\mathbf{h}_{k}=[h_1,h_2,\ldots,h_{N_\mathrm{t}}]$ is an $1 \times N_\mathrm{t}$ vector of complex channel coefficients between the $k$-th antenna of the receiver and the $N_\mathrm{t}$ antennas of the transmitter, $\mathbf{x} = [x_\mathrm{1},x_\mathrm{2},\ldots,x_{N_\mathrm{t}}]^T$ is an $N_\mathrm{t} \times 1$ vector of transmitted precoded complex symbols and $n_{k}$ is the independent identically distributed (i.i.d) zero mean Additive White Gaussian Noise (AWGN) at the $k$-th receiving antenna. The matrix form of (\ref{chan1}) can be expressed as
\begin{equation}
\mathbf{y} = \mathbf{H} \mathbf{x} + \mathbf{n}, \label{chan2}
\end{equation}
where $\mathbf{y} = [y_\mathrm{1},y_\mathrm{2},\ldots,y_{N_\mathrm{r}}]^T$ is $N_\mathrm{r} \times 1$ vector of received signals, $\mathbf{H} = [\mathbf{h}^T_\mathrm{1},\mathbf{h}^T_\mathrm{2},\ldots,\mathbf{h}^T_{N_\mathrm{r}}]^T$ is $N_\mathrm{r} \times N_\mathrm{t}$ channel matrix and $\mathbf{n} = [n_\mathrm{1},n_\mathrm{2},\ldots,n_{N_\mathrm{r}}]^T$ is $N_\mathrm{r} \times 1$ vector of noise variance at the receive antennas. In Constructive Interference Zero Forcing (CIZF) scheme, the cross-interference that is constructive (i.e. increasing symbol power more than required) is not suppressed. Thus, the preserved constructive interference increases the received symbol and decreases the total consumed power at the transmitter \cite{4801492}. In the presented symbol-level optimization technique the problem is initially derived to find optimal solutions ($\mathbf{u}$) to increase amplitude of unprecoded symbols ($\mathbf{s}$) so the total power of the precoded symbols ($\mathbf{x}$) decreases. 
Therefor, we define the transmitted signal $\mathbf{x}$ as
\begin{equation}
\mathbf{x} = \mathbf{W}_\mathrm{ZF} ({\Gamma} \circ \mathbf{s} + \mathbf{u}), \label{chan3}
\end{equation}
where $\mathbf{W}_\mathrm{ZF} = \mathbf{H}^{H} \cdot (\mathbf{H} \cdot \mathbf{H}^{H})^{-1}$ is channel ZF precoding matrix, $\mathbf{s}=[s_{1}, s_{2},\ldots,s_{N_\mathrm{r}}]$ is $N_\mathrm{r} \times 1$ complex vector of the intended symbols with instantaneous unit energy for each receive antenna at the receiver, $\mathbf{u}=[u_1,u_2,\ldots,u_{N_\mathrm{r}}]^T$ is ${N_\mathrm{r}} \times 1$ vector of complex magnitudes of optimal solutions, ${{\Gamma}} = [\sqrt{\gamma_\mathrm{1}},\sqrt{\gamma_\mathrm{2}},$ $\ldots,\sqrt{\gamma_{k}}]^T$ is vector of SNR constraints of symbols at the receiver.
In case of noise variance is equal to zero ($\mathbf{n}=0$) and number of transmit antennas is equal or greater than number of receive antennas ($N_\mathrm{t} \geq N_\mathrm{r}$), the received signal is
\begin{equation}
\mathbf{y} = \mathbf{H}\mathbf{x} = \mathbf{H} \mathbf{W}_\mathrm{ZF} ({\Gamma} \circ \mathbf{s}+\mathbf{u}) = \mathbf{I}_{N_\mathrm{r} \times N_\mathrm{r}} ({\Gamma} \circ \mathbf{s}+\mathbf{u}) = {\Gamma} \circ \mathbf{s} + \mathbf{u}, \label{chan6}
\end{equation}
where $\mathbf{I}_{N_\mathrm{r} \times N_\mathrm{r}}=\mathbf{H}\mathbf{W}_\mathrm{ZF}$. The design of the optimized precoder considers minimizing the total power of the precoded symbols so that the signal received by the $k$-th antenna at the receiver satisfies the thresholds of in-phase and quadrature-phase levels ($\Gamma$). Thereby, we define the optimal precoder design problem such as
\begin{equation}
\label{chan8}
\begin{aligned}
&\min_{\mathbf{x}} &&\lVert \mathbf{x} \rVert_2\\
&\text{subject to} && \mathrm{Re}(\mathbf{h}_{k}\mathbf{x}) \geq \mathrm{Re}(\sqrt{\gamma_{k}}s_{k}), \\ 
&&& \mathrm{Im}(\mathbf{h}_{k}\mathbf{x}) \geq \mathrm{Im}(\sqrt{\gamma_{k}}s_{k}),
\end{aligned}
\end{equation}
for $k = 1 \ldots N_\mathrm{r}$. The amplitude constrains defined in (\ref{chan8}) are only valid if symbol $s_{k}$ is located in the first quadrant of the complex plane. For the optimization problem to hold for symbols from all the quadrants, we have to take into account the sign of in-phase and quadrature-phase values of the symbols. This is done my multiplying the left and right parts of the constants by $\mathrm{Re}(s_{k})/|\mathrm{Re}(s_{k})|$ and $\mathrm{Im}(s_{k})/|\mathrm{Im}(s_{k})|$ accordingly
\begin{equation}
\label{chan8a}
\begin{aligned}
&\min_{\mathbf{x}} &&\lVert \mathbf{x} \rVert_2\\
&\text{subject to} && (\mathrm{Re}(s_{k})/|\mathrm{Re}(s_{k})|) \mathrm{Re}(\mathbf{h}_{k}\mathbf{x}) \geq \mathrm{Re}(\sqrt{\gamma}s_{k}) (\mathrm{Re}(s_{k})/|\mathrm{Re}(s_{k})|), \\ 
&&& (\mathrm{Im}(s_{k})/|\mathrm{Im}(s_{k})|)\mathrm{Im}(\mathbf{h}_{k}\mathbf{x}) \geq \mathrm{Im}(\sqrt{\gamma}s_{k})(\mathrm{Im}(s_{k})/|\mathrm{Im}(s_{k})|).
\end{aligned}
\end{equation}
The problem can be rewritten into the matrix form as
\begin{equation}
\label{chan9}
\begin{aligned}
&\min_{\mathbf{x}} &&\lVert \mathbf{x} \rVert_2\\
&\text{subject to} && \mathbf{b}_{\mathsf{r}} \circ \mathrm{Re}(\mathbf{H}\mathbf{x}) \geq \mathrm{Re}({\Gamma}\circ\mathbf{s}) \circ \mathbf{b}_{\mathsf{r}}, \\ 
&&&\mathbf{b}_{\mathsf{i}} \circ \mathrm{Im}(\mathbf{H}\mathbf{x}) \geq \mathrm{Im}({\Gamma}\circ\mathbf{s}) \circ \mathbf{b}_{\mathsf{i}},
\end{aligned}
\end{equation}
where $\mathbf{b}_{\mathsf{r}}=[\mathrm{Re}(s_{1})/|\mathrm{Re}(s_{1})|, \mathrm{Re}(s_{2})/|\mathrm{Re}(s_{2})|, \ldots, \mathrm{Re}(s_{N_{r}})/|\mathrm{Re}(s_{N_{r}})|]^T$ and $\mathbf{b}_{\mathsf{i}}=[\mathrm{Im}(s_{1})/|\mathrm{Im}(s_{1})|, \mathrm{Im}(s_{2})/|\mathrm{Im}(s_{2})|, \ldots, \mathrm{Im}(s_{N_{r}})/|\mathrm{Im}(s_{N_{r}})|]^T$. By inserting (\ref{chan3}) into (\ref{chan9}) we get
\begin{equation}
\label{chan11}
\begin{aligned}
&\min_{\mathbf{u}} &&\lVert \mathbf{W}_\mathrm{ZF} ({\Gamma}\circ\mathbf{s} + \mathbf{u}) \rVert_2\\
&\text{subject to} && \mathbf{b}_{\mathsf{r}}\circ\mathrm{Re}({\Gamma}\circ\mathbf{s} + \mathbf{u}) \geq \mathrm{Re}({\Gamma}\circ\mathbf{s})\circ \mathbf{b}_{\mathsf{r}}, \\ 
&&&\mathbf{b}_{\mathsf{i}} \circ\mathrm{Im}({\Gamma}\circ\mathbf{s} + \mathbf{u}) \geq \mathrm{Im}({\Gamma}\circ\mathbf{s})\circ \mathbf{b}_{\mathsf{i}}.
\end{aligned}
\end{equation}
Applying linear algebra operations on the constraints of (\ref{chan11}) the problem yields
\begin{equation}
\label{chan12}
\begin{aligned}
&\min_{\mathbf{u}} &&\lVert \mathbf{W}_\mathrm{ZF} ({\Gamma}\circ\mathbf{s}) + \mathbf{W}_\mathrm{ZF}\mathbf{u} \rVert_2\\
&\text{subject to} && \mathbf{b}_{\mathsf{r}}\circ\mathrm{Re}(\mathbf{u}) \geq \mathbf{0}, \\ 
&&&\mathbf{b}_{\mathsf{i}}\circ\mathrm{Im}(\mathbf{u}) \geq \mathbf{0}.
\end{aligned}
\end{equation}
The following step towards problem unification is to remove the real and imaginary valued parts from (\ref{chan12}). This can be done by expressing  $\mathbf{W}_\mathrm{ZF} = \mathrm{Re}(\mathbf{W}_\mathrm{ZF}) + i\mathrm{Im}(\mathbf{W}_\mathrm{ZF})$, $\mathbf{s} = \mathrm{Re}(\mathbf{s}) + i\mathrm{Im}(\mathbf{s})$ and $\mathbf{u} = \mathrm{Re}(\mathbf{u}) + i\mathrm{Im}(\mathbf{u})$ and separating the real and imaginary parts of $\mathbf{W}_\mathrm{ZF}({\Gamma}\circ\mathbf{s})$ and $\mathbf{W}_\mathrm{ZF}\mathbf{u}$ as
\begin{equation}
\label{chan13}
\begin{aligned}
\begin{split}
\mathbf{W}_\mathrm{ZF}(\Gamma\circ\mathbf{s}) = \mathrm{Re}(\mathbf{W}_\mathrm{ZF})\mathrm{Re}(\Gamma\circ\mathbf{s}) - \mathrm{Im}(\mathbf{W}_\mathrm{ZF})\mathrm{Im}(\Gamma\circ\mathbf{s}) \\+i[\mathrm{Re}(\mathbf{W}_\mathrm{ZF})\mathrm{Im}(\Gamma\circ\mathbf{s})+\mathrm{Im}(\mathbf{W}_\mathrm{ZF})\mathrm{Re}(\Gamma\circ\mathbf{s})],
\end{split}
\end{aligned}
\end{equation}
\begin{equation}
\label{chan14}
\begin{aligned}
\begin{split}
\mathbf{W}_\mathrm{ZF}\mathbf{u} = \mathrm{Re}(\mathbf{W}_\mathrm{ZF})\mathrm{Re}(\mathbf{u}) - \mathrm{Im}(\mathbf{W}_\mathrm{ZF})\mathrm{Im}(\mathbf{u}) \\+i[\mathrm{Re}(\mathbf{W}_\mathrm{ZF})\mathrm{Im}(\mathbf{u})+\mathrm{Im}(\mathbf{W}_\mathrm{ZF})\mathrm{Re}(\mathbf{u})].
\end{split}
\end{aligned}
\end{equation}
From (\ref{chan13}) and (\ref{chan14}), we derive
\begin{subequations}
\begin{align}
\mathbf{W}_\mathrm{ZF}(\Gamma\circ\mathbf{s}) & = \mathbf{W}_\mathrm{ZF1}(\overline{\Gamma}\circ\bar{\mathbf{s}}) +i\mathbf{W}_\mathrm{ZF2}(\overline{\Gamma}\circ\bar{\mathbf{s}}), \label{chan15a} \\
\mathbf{W}_\mathrm{ZF}\mathbf{u} & = \mathbf{W}_\mathrm{ZF1}\bar{\mathbf{u}}+i\mathbf{W}_\mathrm{ZF2}\bar{\mathbf{u}}, \label{chan15b}
\end{align}
\end{subequations}
where $\mathbf{W}_\mathrm{ZF1}= [\mathrm{Re}(\mathbf{W}_\mathrm{ZF}), - \mathrm{Im}(\mathbf{W}_\mathrm{ZF})]$, $\mathbf{W}_\mathrm{ZF2}= [\mathrm{Im}(\mathbf{W}_\mathrm{ZF}), \mathrm{Re}(\mathbf{W}_\mathrm{ZF})]$, $\bar{\mathbf{s}} = [\mathrm{Re}(\mathbf{s}^T),$ $ \mathrm{Im}(\mathbf{s}^T)]^T$, $\bar{\mathbf{u}} = [\mathrm{Re}(\mathbf{u}^T), \mathrm{Im}(\mathbf{u}^T)]^T$, and $\overline{\Gamma} = [{\Gamma}^T, {\Gamma}^T]^T$.
By taking into consideration that the Euclidean norm of complex vector $\tilde{z} = [\tilde{z_1}, \tilde{z_2}, \ldots, \tilde{z_i}]$, where $\tilde{z_i} = a_i+ib_i$, is equivalent to the Euclidean norm of a real vector $z = [z_1, z_2, \ldots, z_i]$, where $z_i = [a_i, b_i]$, it is valid to define the following equality
\begin{subequations}
\begin{align}
\lVert\mathbf{x}\rVert_2 & = \lVert\bar{\mathbf{x}}\rVert_2, \label{chan16a} \\
\lVert \mathbf{W}_\mathrm{ZF} ({\Gamma}\circ\mathbf{s}) + \mathbf{W}_\mathrm{ZF}\mathbf{u} \rVert_2 & = \lVert \overline{\mathbf{W}}_\mathrm{ZF} (\overline{\Gamma}\circ\bar{\mathbf{s}}) + \overline{\mathbf{W}}_\mathrm{ZF}\bar{\mathbf{u}} \rVert_2, \label{chan16b}
\end{align}
\end{subequations}
where $\bar{\mathbf{x}} = [\mathrm{Re}(\mathbf{x}^T),\mathrm{Im}(\mathbf{x}^T)]^T$ and $\overline{\mathbf{W}}_\mathrm{ZF} = [\mathbf{W}_\mathrm{ZF1}^T, \mathbf{W}_\mathrm{ZF2}^T]^T$. By inserting (\ref{chan16b}) to (\ref{chan12}) the optimization problem turns into
\begin{equation}
\label{chan17}
\begin{aligned}
&\min_{\bar{\mathbf{u}}} &&\lVert \overline{\mathbf{W}}_\mathrm{ZF} (\overline{\Gamma}\circ\bar{\mathbf{s}}) + \overline{\mathbf{W}}_\mathrm{ZF}\bar{\mathbf{u}} \rVert_2\\
&\text{subject to} &&\mathbf{b} \circ \bar{\mathbf{u}} \geq 0 \\
\end{aligned}
\end{equation}
where $\mathbf{b} = [\mathbf{b}_{\mathsf{r}}^T, \mathbf{b}_{\mathsf{i}}^T]^T$, $\mathrm{Re}(\bar{\mathbf{u}})=\bar{\mathbf{u}}$ and $\mathrm{Im}(\bar{\mathbf{u}})=0$.
To solve (\ref{chan17}), we without lost of problem definition can change the equality of (\ref{chan16b}) as
\begin{equation}
\overline{\mathbf{W}}_\mathrm{ZF} (\overline{\Gamma}\circ\bar{\mathbf{s}}) + \overline{\mathbf{W}}_\mathrm{ZF}\bar{\mathbf{u}} = \overline{\mathbf{W}}_\mathrm{ZF}\mathbf{B} (\overline{\Gamma}\circ(\bar{\mathbf{s}}\circ\mathbf{b})) + \overline{\mathbf{W}}_\mathrm{ZF}\mathbf{B}(\bar{\mathbf{u}}\circ\mathbf{b}) \label{chan18}
\end{equation}
where $\mathbf{B}$ - diagonal matrix where elements of the vector $\mathbf{b}$ are its diagonal entries. The element-wise multiplication $(\bar{\mathbf{s}}\circ\mathbf{b})$ rotates symbols into the first quadrant of the complex plane as it is shown in Fig. \ref{fig-1}. The modified precoding matrix $\overline{\mathbf{W}}_\mathrm{ZF}\cdot\mathbf{B}$ accordingly accounts the symbol rotation so the precoded symbols $\bar{\mathbf{x}}$ does not change. By replacing $\overline{\mathbf{W}}_\mathrm{ZF}\cdot\mathbf{B} = \widetilde{\mathbf{W}}_\mathrm{ZF}$, $\bar{\mathbf{s}}\circ\mathbf{b} = \tilde{\mathbf{s}}$ and $\bar{\mathbf{u}}\circ\mathbf{b} = \tilde{\mathbf{u}}$ in the (\ref{chan18}) and inserting it back to (\ref{chan17}) the problem becomes
\begin{equation}
\label{chan21}
\begin{aligned}
&\min_{\tilde{\mathbf{u}}} &&\lVert\widetilde{\mathbf{W}}_\mathrm{ZF}\tilde{\mathbf{u}}-\mathbf{d} \rVert_2\\
&\text{subject to} &&\tilde{\mathbf{u}} \geq 0,
\end{aligned}
\end{equation}
where $\mathbf{d} = - \widetilde{\mathbf{W}}_\mathrm{ZF}(\Gamma\circ\tilde{\mathbf{s}})$. The problem (\ref{chan21}) is NNLS optimization problem which can be solved using fast NNLS algorithm \cite{CEM:CEM483}. The optimized linear precoder is therefore defined as
\begin{equation}
\begin{aligned}
\bar{\mathbf{x}} & = \widetilde{\mathbf{W}}_\mathrm{ZF}(\overline{\Gamma}\circ\tilde{\mathbf{s}}+\tilde{\mathbf{u}}). \label{chan22}
\end{aligned}
\end{equation}
After calculating new values of precoded symbols using (\ref{chan22}), the real and imaginary parts of the precoded signal $x_i$ for $i = 1 \ldots N_\mathrm{t}$ can be extracted as
\begin{equation}
\begin{aligned}
x_{i} & = \bar{x}_{i} + i  \bar{x}_{i+ N_\mathrm{t}}. \label{chan23}
\end{aligned}
\end{equation}

\section{Adapting the optimal solution to M-PSK and M-APSK constellations} \label{optcondesign}
The optimization problem defined in (\ref{chan21}) is a straightforward solution for QPSK constellation. The solutions $\tilde{\mathbf{u}}$ are provided in a wide region as shown in Fig. \ref{fig-1}. This will result in phase rotation of the optimized symbols after adding the $\tilde{\mathbf{u}}$ to $\tilde{\mathbf{s}}$ in (\ref{chan22}). In case of QPSK constellation the optimized symbol will stay in right detection region as there is only one symbol per quadrant of the complex plane. In case of M-PSK (M $>$ 4) and M-APSK constellations, this may result in symbols out of their correct detection regions. To avoid the symbols going out of their detection regions we go through a post-processing of the vector $\tilde{\mathbf{u}}$ to verify the phase of the optimized symbols according to constellation scheme under consideration.

\subsection{Detection region verification for M-PSK constellation}
\label{optcondesign_psk}

In this section we devise a routine of a detection region verification for M-PSK symbols. From the Fig. \ref{fig-1} we see, that point of optimal solution can be in three different regions on the complex plane: in case of (1.) the point is below detection region defined in M-PSK constellation, (2.) - above it, and  (3.) - the point is inside the detection region. If either of first two cases appear, the optimal solution has to be decreased until the optimized symbol is again on the edge of the detection region. In case of (1.) the In-phase part of the optimal point has to be reduced, while in case if (2.) the Quadrature part of optimal solution. To check if the optimized symbols are inside detection region, we verify the ratio of Quadrature-phase ($\tilde{u}_{k+N_{\mathrm{r}}}$) and In-phase ($\tilde{u}_{k}$) parts of optimal solutions for each symbol for $k = 1 \ldots N_\mathrm{r}$. We define a desirable threshold angle of the phase of an optimized symbol as $\theta_\mathrm{0} = \pi/M$, where $M$ - the order of M-PSK modulation scheme and $\theta_{\mathrm{s}_{k}}$ - the argument of the symbol vector $\mathrm{s}_{{k}}$. The ratio ($\tilde{u}_{k+N_{\mathrm{r}}}/\tilde{u}_{k}$) has to be greater or equal than $\tan (\theta_{\mathrm{s}_{{k}}} - \theta_\mathrm{0})$ and less or equal than $\tan (\theta_{\mathrm{s}_{{k}}}  + \theta_\mathrm{0})$ for the optimal point to be inside the M-PSK detection region. By solving the following equalities we determine if the optimal solution occurs in either case (1.) or (2.) 
\begin{subequations}
\begin{align}
{\delta_r}_k &=\tilde{u}_\mathrm{k+N_{\mathrm{r}}}/\tilde{u}_\mathrm{k}-\tan (\theta_{\mathrm{s}_{\mathrm{k}}} - \theta_\mathrm{0}), \label{chan27a}\\
{\delta_i}_k &= {\tan (\theta_{\mathrm{s}_{\mathrm{k}}} + \theta_\mathrm{0})} -\tilde{u}_\mathrm{k+N_{\mathrm{r}}}/\tilde{u}_\mathrm{k}. \label{chan27b}
\end{align}
\end{subequations}
If ${\delta_r}_k$ of (\ref{chan27a}) is less than zero, the point of optimal solution is in case (1.). If ${\delta_i}_k$ of (\ref{chan27b}) is less than zero - in case (2.). To resolve the issue of case (1.) we set ${\delta_r}_k$ to zero and solve the equation for In-phase value ($\tilde{u}_\mathrm{k}$). For the case (2.) we set ${\delta_i}_k$ equal to zero and solve the equation for $\tilde{u}_\mathrm{k+N_{\mathrm{r}}}$.
We denote the corrected optimization solutions with respect to detection region of the symbols $\tilde{\mathbf{u}}$ as
\begin{subequations}
\begin{align}
\label{chan28a}
  \tilde{u}_{k} &=
  \begin{cases}
    \bar{u}_\mathrm{k+N_{\mathrm{r}}}/\tan (\theta_{\mathrm{s}_{\mathrm{k}}} - \theta_\mathrm{0}) & \text{if ${\delta_r}_k < 0$, for $k = 1\ldots N_{\mathrm{r}}$,}\\
    \bar{u}_{k} & \text{if ${\delta_r}_k \geq 0$, for $k = 1\ldots N_{\mathrm{r}}$,}
  \end{cases}\\
    \label{chan28b}
  \tilde{u}_{k+N_{\mathrm{r}}} &=
  \begin{cases}
    \bar{u}_\mathrm{k}\cdot \tan (\theta_{\mathrm{s}_{\mathrm{k}}} + \theta_\mathrm{0}) & \text{if ${\delta_i}_k < 0$, for $k = 1\ldots N_{\mathrm{r}}$,}\\   
    \bar{u}_{k+N_{\mathrm{r}}}   & \text{if ${\delta_i}_k \geq 0$, for $k = 1\ldots N_{\mathrm{r}}$}.
  \end{cases}
\end{align}  
\end{subequations}


\subsection{Detection region verification for M-APSK constellation}
\label{optcondesign_apsk}
The optimization in the M-APSK constellation is applicable to symbols on the top level only. Thus, all the optimal solutions for symbols, which are not from the top level of the constellation have to be reset to zero. Therefor, we go through an additional processing of $\tilde{u}_{k}$ defined in (\ref{chan28a}) and (\ref{chan28b}) as follows
\begin{equation}
\label{chan30a}
  \tilde{u}_{k} =
  \begin{cases}
    0 & \text{if $|s_\mathrm{k}|^2<P_{t}$, for $k = 1\ldots N_{\mathrm{r}}$},\\
    \tilde{u}_{k} & \text{if $|s_\mathrm{k}|^2\geq P_{t}$, for $k = 1\ldots N_{\mathrm{r}}$},
  \end{cases}
\end{equation}
\begin{equation}
\label{chan30b}
  \tilde{u}_{k+N_{\mathrm{r}}} =
  \begin{cases}
    0 & \text{if $|s_\mathrm{k}|^2<P_{t}$, for $k = 1\ldots N_{\mathrm{r}}$},\\
    \tilde{u}_{k+N_{\mathrm{r}}} & \text{if $|s_\mathrm{k}|^2\geq P_{t}$, for $k = 1\ldots N_{\mathrm{r}}$},
  \end{cases}
\end{equation}
where $P_{t}$ - is modulus of complex value of symbol on top level of the constellation.

Regardless of the processing introduced in sections \ref{optcondesign_psk} and \ref{optcondesign_apsk}, the definition and complexity of the initial optimization problem (\ref{chan21}) remains the same for all the types of symbol constellations.


\section{Benchmark results} \label{results}

\begin{figure}[!tbp]
  \centering
  \begin{minipage}[b]{0.45\textwidth}
    \includegraphics[width=1.1\textwidth]{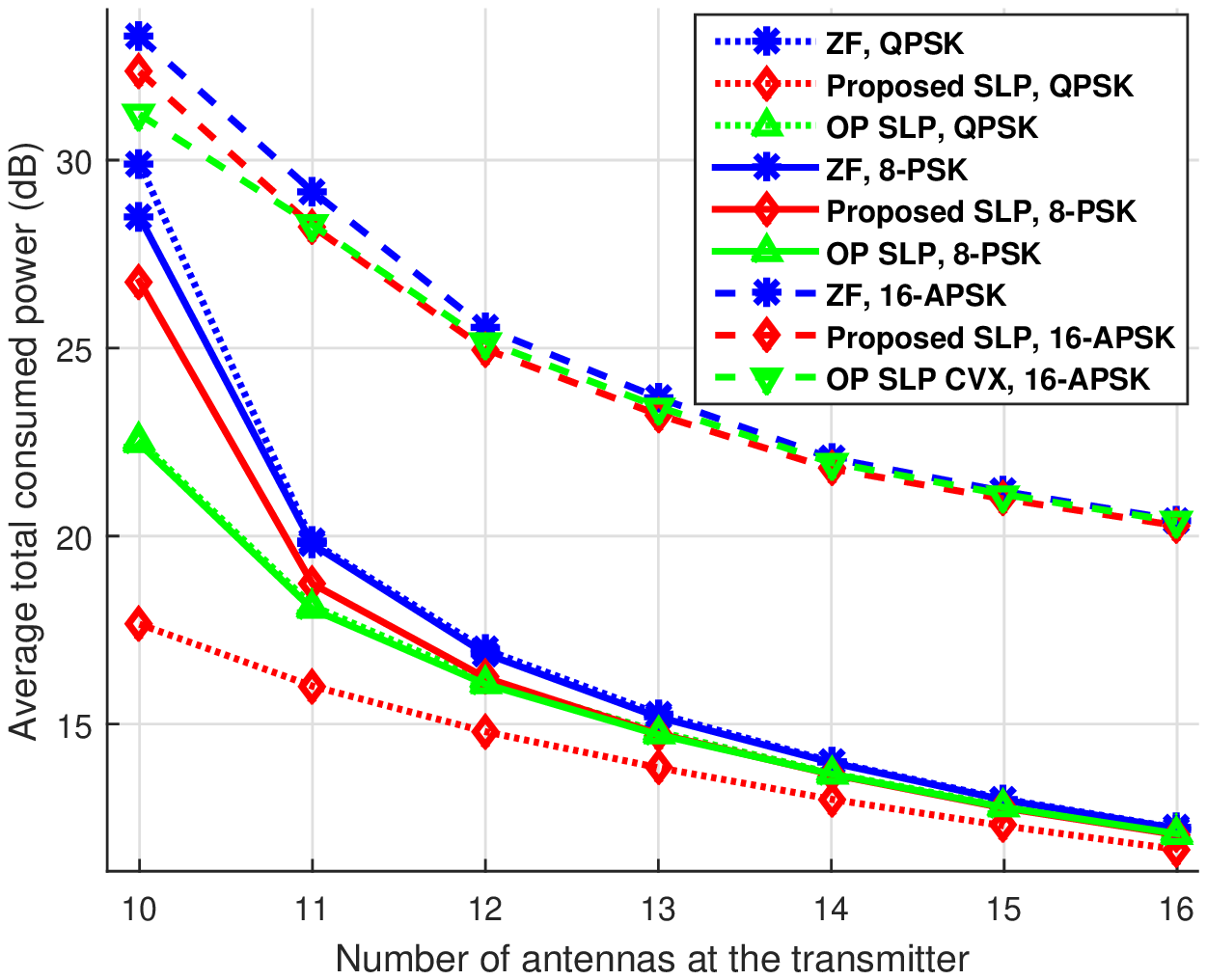}
    \caption{Average total power.}
    \label{fig-4}
  \end{minipage}
  \hfill
  \begin{minipage}[b]{0.45\textwidth}
    \includegraphics[width=1.1\textwidth]{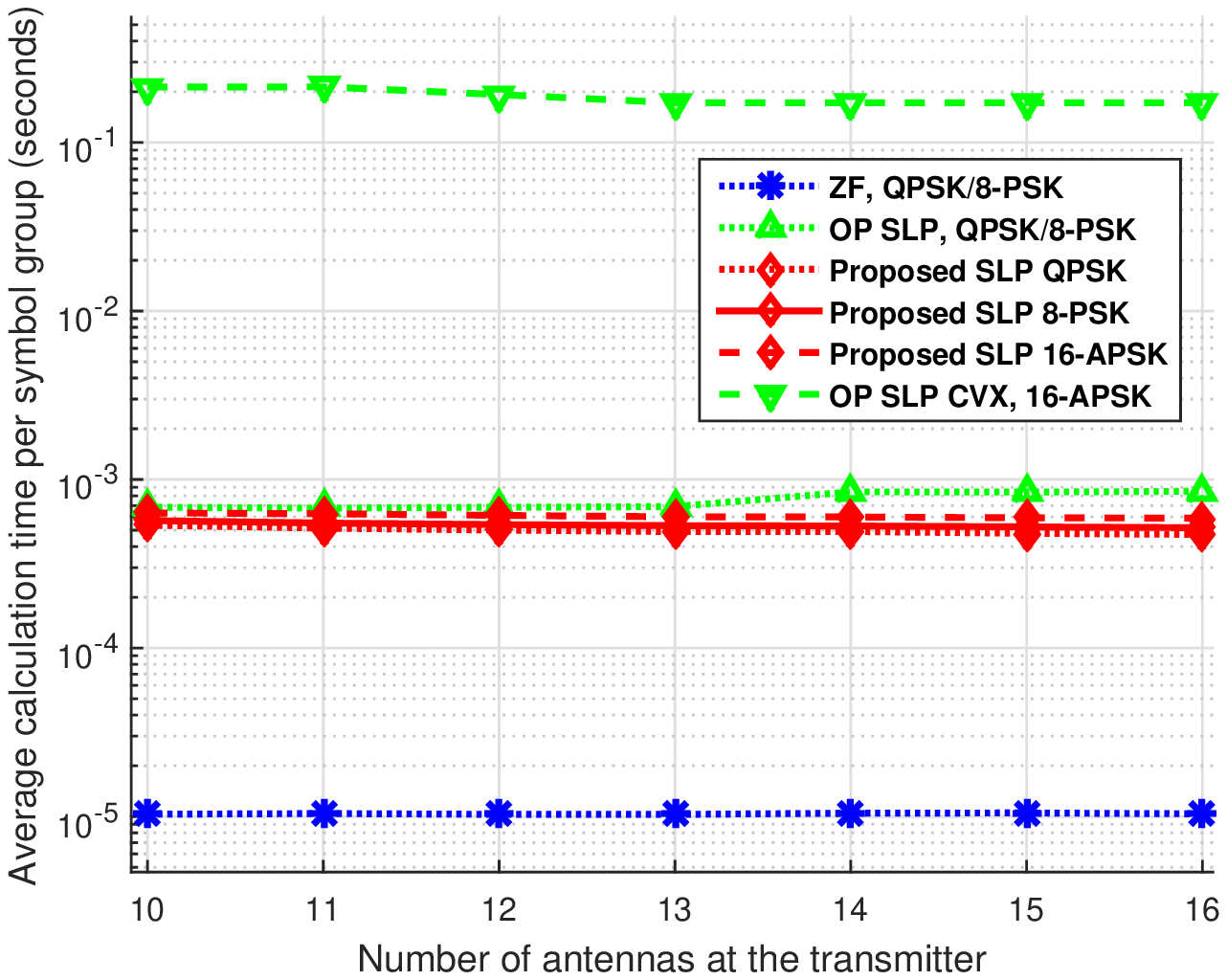}
    \caption{Average processing time.}
    \label{fig-5}
  \end{minipage}
  \newline {Results of benchmarks for QPSK, 8-PSK and 16-APSK constellations,\\ $N_\mathrm{r}=10$, {$\Gamma$ = 10 dB}.}
\end{figure}

To compare the performance of the proposed technique with the literature, we use averaged results over total consumed power and processing time per set of symbols as performance metrics. We use ZF precoder \cite{4840357}, optimal symbol-level precoder of \cite{7544467} and an symbol-level precoding for multi-level constellations of \cite{AlodehCO16} as the benchmark schemes. These techniques are denoted by "ZF", "OP SLP", and "OP SLP CVX". The proposed symbol-level precoder is denoted "Proposed SLP". User symbols and channel matrix are generated using random values with normal distribution. The total power consumption benchmark is shown on Fig. \ref{fig-4}. The processing time benchmark is shown on Fig. \ref{fig-5}.

In the case of QPSK constellation the proposed technique provides the lowest total power consumption for all given $N_\mathrm{t}$. The minimal power is achieved at $N_\mathrm{t}$ = $N_\mathrm{r}$, which is 9 dB below the benchmark of the ZF precoder.

In case of the 8-PSK constellation and $N_\mathrm{t} = 10$ the power consumption is 2 dB below the ZF precoder results. The optimal symbol-level precoder of \cite{7544467} in the same conditions provides results of 4.9 dB below the ZF, which is considerable better (2.9 dB lower) result comparing to the proposed method. If the parameter $N_\mathrm{t}$ is larger than $N_\mathrm{r}$ the gap between proposed method and \cite{7544467} shrinks to 0.8 dB. 
The processing time of the proposed procoder is shorter than other benchmark precoders. The extra processing routine, which was introduced in the section \ref{optcondesign}, has a very minor impact on the processing time. The proposed and technique of \cite{7544467} are both NNLS optimization problems. But the latest in additional depends on a performance of a singular value decomposition (SVD) algorithm. 

In case of the 16-APSK constellation the performance of the proposed precoder is similar to other cases. The proposed precoder achieves lower power consumption than ZF precoder, but higher than the precoder of \cite{AlodehCO16} for $N_\mathrm{t} = N_\mathrm{r}$ case. The performance gap between the two methods drastically reduces then the number of transmit antennas is greater than receive antennas ($N_\mathrm{t} > N_\mathrm{r}$). The proposed method performs at the remarkable shorter processing times per set of symbols. The proposed precoder designed as NNLS optimization problem even in case of multi-level constellations, which results in remarkable shorter processing time per set of symbols.

In real implementation of MIMO precoding systems the processing time of symbols is crucial parameter especially in the high-throughput mode. The proposed technique provides faster processing time per set of symbols for all constellation types in the benchmark. Minimization of the consumed power is not optimal for constellations of high-order symbol modulation ($\mathbf{M} > 4$), although, the difference between the presented precoder and the benchmark techniques becomes negligible for $N_\mathrm{t} > N_\mathrm{r}$.

%
%

\section{Conclusions} \label{conclusions}
In this work, we derived low complexity symbol-level power optimization technique for linear precoding systems. The presented precoding technique designs the precoded symbols with total power lower than ZF precoder, while sustaining the minimal required SNR at the receiver. It achieves the shortest processing time to design optimized precoded symbols among other optimal symbol-level precoding techniques under the benchmark. The precoder utilizes NNLS optimization design for all types of symbol constellations, which simplifies the implementation and the complexity of the technique. Hence, the proposed algorithm to design power minimization symbol-level precoder is a good candidate for integration into realistic precoded transmitter.

\section*{Acknowledgement}
This research was supported by Luxembourg National Research Fund grant for AFR-PPP project "End-to-end Signal Processing Algorithms for Precoded Satellite Communications" in collaboration with FNR "SERENADE" project.

\bibliographystyle{IEEEtran}
\bibliography{mimo}

\end{document}